# Colossal barocaloric effects in plastic crystals

Bing Li[1]*, Yukinobu Kawakita[2], Hui Wang[3], Tatsuya Kikuchi[2], Dehong Yu[4], Shinichiro Yano[5], Weijun Ren[1], Kenji Nakajima[2], and Zhidong Zhang[1]

[1]Institute of Metal Research (IMR), Chinese Academy of Sciences, Shenyang 110016, China

[2]J-PARC Center, Japan Atomic Energy Agency, Tokai, Ibaraki 319-1195, Japan.

[3]Department of Physics and Astronomy, University of California, Irvine, California 92697, USA.

[4]Australian Nuclear Science and Technology Organization (ANSTO), Lucas Heights, New South Wales 2232, Australia.

[5]Neutron Group, National Synchrotron Radiation Research Center, Hsinchu 30077, Taiwan.

*Correspondence to: bingli@imr.ac.cn (B.L.).

**Abstract**: Solid-state refrigeration technology based on caloric effects are promising to replace the currently used vapor compression cycles. However, their application is restricted due to limited performances of caloric materials. Here, we have identified colossal barocaloric effects (CBCEs) in a class of disordered solids called plastic crystals. The obtained entropy changes are about 380 J $kg^{-1}$ $K^{-1}$ in the representative neopentylglycol around room temperature. Inelastic neutron scattering reveals that the CBCEs in plastic crystals are attributed to the combination of the vast molecular orientational disorder, giant compressibility and high anharmonic lattice dynamics. Our study establishes the microscopic scenario for CBCEs in plastic crystals and paves a new route to the next-generation solid-state refrigeration technology.

Refrigeration are of vital importance for the modern society, such as for food storage and air conditioning (*1*). Nowadays, 25-30% world's electricity is consumed for variety of purposes of refrigeration, mostly involving the conventional vapor compression cycle technology (*2*). Their working materials are subjected to the growing environmental concerns. As a promising replacement, the refrigeration technology based on solid-state caloric effects has been attracting extensive attention in recent decades (*3*).

Caloric effects arise due to the fact that the disorder of a degree of freedom in solids can be effectively suppressed by an external field around an order-to-disorder transition. During such a process, isothermal entropy changes and adiabatic temperature changes are detected, which are the most important assessments for a caloric-effect material. In the vicinity of a ferromagnetic-to-paramagnetic transition, an applied magnetic field can effectively suppress the disorder of magnetic moments, which in turn account for the entropy changes, widely known as the magnetocaloric effect (MCE) (*4*). Similarly, an electrocaloric effect (ECE) can be also found near a ferroelectric-to-paraelectric transition with the tuning of electric fields (*5*). In some elastically-active compounds known as ferroelastics, an external compression or tensile stress can also induce a modification of the crystal structure, which brings out the elastocaloric effect (eCE) (*6*). Unlike above three caloric effects that are observed in ferroics, a barocaloric effect (BCE) is not system-selective and, in principle, can be induced at any atomic systems by applying hydrostatic pressure because it is always factored in the free energy of a system. In this sense, BCE is most universal among these solid-state caloric effects.

However, current caloric effects are characteristic of entropy changes about dozens of J $kg^{-1}K^{-1}$, which is far away from the expected. Thus, it is urgent to enhance caloric effects by optimizing working materials, in particular, to explore larger BCE. Here, we have identified that the plastic crystals are very promising next-generation BCE materials with colossal entropy changes under relatively smaller pressures.

Plastic crystals are a class of highly disordered molecular solids (*7*). They are usually referred as to the orientation-disordered and position-ordered molecular solids where the organic molecules orient randomly while their gravity centers form the highly-symmetry lattices like face-center-cubic (FCC) one. With cooling down, the molecules exhibit preferred orientations, which results in a disorder-to-order transition with lattice symmetry-breaking. During the transitions, huge entropy changes are accompanied because of the vast accessible configurations in the disordered state. Owing to this feature, they have been used as solid state thermal energy storage materials (*7*). As shown in Fig. 1a, their enthalpies of the solid-state transition from the ordered state to the plastic phase are significantly larger than the fusion enthalpies (*8-10*), in sharp contrast to the regular solids such as Fe (*11*), Mn (*12*), Ti (*13*) and SnSe (*14*). This means most of the heat stored in plastic crystals are released at this solid-state transition rather than at the melting, which is characterized as so-called two-step melting process. In other words, most of atomic degrees of freedom become liquid-like while only the gravity centers form the lattice to sustain the rigidity. In the atomic dynamics, the plastic crystals show the ultrafast reorientations with the timescales ranging from microseconds to picoseconds (*15*), which resembles the dynamics of molecular liquids. Based on these two merits, we categorize plastic crystals as solid-liquid hybrid crystals, in order to emphasize that these materials are macroscopically solid while microscopically liquid-like.

As another important ingredient, extreme sensitivity to applied pressure is a prerequisite for CBCE, which implies that a tiny pressure can induce a huge response. Plastic crystals are usually soft and easy to be deformed. This is the reason why they are named. This ability originates from the large compressibility that is inversely proportional to the density and square of sound speed. We compare the compressibility of leading plastic crystals with that of regular solids in Fig. 1b (*16-18*). It can be seen that the former is almost two order of magnitudes larger. In addition, according to the Clausius–Clapeyron relation, $\mathrm{d}P/\mathrm{d}T$ is involved when the induced volumetric change is translated into entropy changes. By a simply thermodynamic deduction, one is able to obtain that $\mathrm{d}P/\mathrm{d}T$ is proportional to Grüneisen parameter that is a measure of how anharmonic a system is (*19*).

The combination of the vast disorder, giant compressibility and strong anharmonicity makes plastic crystals the ideal host for BCE. As shown in Fig. 1c, entropy changes of about 629 J $kg^{-1}$ $K^{-1}$ (*16*) can be obtained around 350 K in AMP while 380 J $kg^{-1}$ $K^{-1}$ in NPG around room temperature. The former stands the largest among all known caloric materials (*3-6*, *20-26*). Focusing on the representative NPG, we employ high-resolution quasi-elastic neutrons scattering (QENS) and inelastic neutron scattering (INS) techniques to establish the microscopic scenario for CBCE in plastic crystals.

The molecular configuration of NPG is shown in Fig. 2a (*27*). Five carbon atoms consist of the centered tetrahedron, where two are attached to hydrogen atoms in the methyl group, while another two form the hydroxymethyl group. At room temperature, the NPG molecules are ordered on a monoclinic lattice with space group of $P2_1/c$. Intermolecular hydrogen bonds between O and H atoms bridge the adjacent molecules to form infinite hydrogen-bond ladders along the *a* axis (Fig. S1). Just few kelvin higher, this ordered phase transforms into a FCC lattice at $T_t$ of ~313 K, as evidences in the neutron diffraction patterns where the (011) Bragg peak disappears (Fig. 2b). At the same time, this transition is coupled with the hydrogen atoms since the incoherent scattering intensity shows an upturn at $T_t$ (Fig. 2c).

Now, we move to the dynamic aspects of this compound. In an INS spectrum as a function of energy transfer ($E$), the peak centered at $E = 0$ represents the elastic information from static structure while a few peaks centered at $E \neq 0$ are inelastic scattering originating from excitations such as phonons. In some special systems, there exist diffusive and/or reorientaional motions and hence a broader Lorentzian function is imposed underneath the elastic line, which is called QENS (*28*). Even though QENS is centered at $E = 0$, similar to the elastic line, it exhibits finite linewidth that is inversely related to the typical time scales of the motions. The QENS on a hydrogen-contained molecular system gives rise to direct information on how the molecules move. QENS measurements were performed at selected temperatures near $T_t$ on NPG. Fig. 3a show the contour plots of dynamic structure factor $S(Q,E)$ as a function of momentum transfer ($Q$) and energy transfer ($E$) with incident neutron energy ($E_i$) of 2.6363 meV. The intense stripes centered at $E = 0$ represent the elastic line which contains most of the scattering intensity. Such elastic lines dominate the spectra of 250 and 300 K. At 320 K, just above $T_t$, a less intense signal is found to spread out from $E = 0$ and getting weaker as $E$ increases. This is just the QENS originating from the incoherent scattering intensity of hydrogen atoms. QENS intensity is $Q$ dependent as they are weaker and more diffuse at higher $Q$ regions.

At selected $Q$ positions where the Bragg peaks are absent, two dimensional $S(Q,E)$ data were sliced for the standard QENS fitting. Shown in Fig. 3b and c are the $S(E)$ data sliced for $0.9 \leq Q \leq 1.0$ Å$^{-1}$ at 300 and 320 K, respectively. It can be seen that the 300 K data can be well reproduced by a combination of a delta function, a Lorentzian function and a constant background, which are convoluted with the instrumental resolution. However, an extra Lorentzian function is needed to reproduce the data at 320 K. One is narrower while another is much wider. Their linewidth ($\Gamma$) are 10 times different. Repeating the fitting at other $Q$ positions, we are able to obtain the intensity of each components and $\Gamma$ for Lorentzian function(s). The $Q$ dependence of $\Gamma$ are summarized in Fig. 3c and d. Both modes exhibit a weak $Q$ dependence, indicating the associated motions are highly localized. The temperature dependence of $\Gamma$ for $0.9 \leq Q \leq 1.0$ Å$^{-1}$ is plotted in Fig. 2d. This is fitted to the Arrhenius relation and the obtained $E_a$ is about 47(3) meV.

Based on these results, we introduce the elastic incoherent scattering factor (EISF). EISF is defined as the ratio between elastic intensity and total intensity, whose $Q$ dependence is directly related the nature of the mode, i.e., how the QENS intensity is enhanced as $Q$ increases. As shown in Fig. 3d, EISF at 300 K can be well reproduced based on the three-fold ($C_3$) reorientation modes of hydrogen atoms in the methyl group, since the obtained H-H distance in the fitting is 0.82(4) Å. At 320 K, EISF for both modes are plotted in Fig. 3e. We can see the EISF of the slow mode decays much quicker than that of the fast one. The latter is almost identical to the EISF at 300 K, which can be also well reproduced with a partial $C_3$ model. While, the slow one is close to an isotropic reorientation mode with the radius of 0.85(5) Å. Such a distance approximates the gravity of hydrogen atoms if we simplify the molecule as a pole configuration (Fig. S6). Overall, the hydrogen atoms in the methyl group are subjected to the $C_3$ reorientation in both ordered and disordered phases, while the whole molecule undergoes the isotropic reorientation in the disordered phase. This is supported by the thermal measurement. In a FCC lattice, the $T_d$ and $C_{3v}$ symmetries lead to ten equivalent configurations for a tetrahedron molecule (*29*). As for NPG, the internal degrees of freedom is 6 for the arrangements of two hydroxymethyl groups on the corners of tetrahedron. As a result, the total configurations are 60, which lead to entropy changes $R\ln60$ ($R$ is the gas constant). Compared with this common case, the entropy changes of our system is much larger and isotropic reorientation must to be responsible, in agreement with the QENS results. The activation energy of this isotropic reorientation mode is shown in Fig. 2d.

As we discussed in the beginning, the anharmonic lattice dynamics plays a crucial role in BCE. The INS was employed to investigate the phonons using both AMATEAS and Pelican time-of-flight spectrometers. As shown in Fig. 4a is the contour plot of $S(Q,E)$ at 5 K up to 20 meV. It can be seen that a series of bands centered at $E \neq 0$, which represent optical phonons. The $Q$-sliced data at $3 \leq Q \leq 4$ Å$^{-1}$ is plotted in Fig. 4b. These phonons appear as well-defined peaks. The DFT calculation is used to interpret these modes. We can see most of them are identical with each other. As the temperature increase, the phonon peaks are gradually broadened and become completely undistinguished above $T_t$ due to the molecular reorientation disorder (Fig. S7). Focusing on the mode at 12.7 meV, which is related to stretching motions, we track the temperature dependence, as shown in the inset of Fig. 4b. It is clear that the mode become significantly softened with increasing temperature, indicative of giant anharmonicity.

Plastic crystals are emergent playground for solid-state refrigeration. For a list of plastic crystals, please refer to Table S2. Three crucial issues would be the focus in the future, as follows:

a. The BCE of plastic crystals is directly related to the orientational disorder of molecules. Thus, it is expected that a plastic crystal with more internal degrees of freedom is promising to show even higher BCE, because of more degenerate configurations. In fact, the entropy changes are determined by $R\ln\Omega$, where $\Omega$ is the numbers of configurations. This is different from the magnetic entropy changes that is determined as $R\ln(2J+1)$, where $J$ is the quantum number. Thus, the maximum magnetic entropy changes are limited by the maximum of $J = 8$ (for Holmium trivalent ions).

b. The working span of plastic crystals is highly tunable by making solid solution between different plastic crystals, which is necessary for different purposes of refrigeration. For example, mixture of NPG and AMP can cover a temperature from ~ 310 to 350 K (*8*).

c. The thermal hysteresis is intrinsic for a first-order transition, as observed for all caloric-effect materials. Towards the practical applications, the thermal hysteresis has to be tamed. According to the knowledge accumulated in previous research, it is promising to use alloying (*8*) and microstructure engineering (*30*). Specially, a plastic crystal has another possibility. The orientational and positional degrees of freedom are internally coupled and dedicate tuning of their activation energies gives rise to a crossover from first-order to second-order (*31*).

**References:**

1. The Importance of Energy Efficiency in the Refrigeration, Air-conditioning and Heat Pump Sectors. United Nations Environmental Programme, May 2018.

2. Savings and benefits of global regulations for energy efficient products, Ecofys, 2015.
3. X. Moya et al., *Nat. Mater.* **13**, 439 (2014).
4. V. Franco et al., *Prog. Mater. Sci.* **93**, 112 (2018).
5. J. F. Scott, *Annu. Rev. Mater. Res.* **41**, 229 (2011).
6. L. Mañosa et al., *Adv. Mater.* **29**, 1603607 (2017).
7. L. A. K. Staveley, *Annu. Rev. Phys. Chem.* **13**, 351 (1962).
8. D. Chandra et al., *J. Less Common. Met.* **168**, 159 (1991).
9. E. Murrill et al., *Thermochim. Acta* **1**, 239 (1970).
10. J. L. Tamarit et al., *J. Phys. Condens. Matter* **9**, 5469 (1997).
11. P. D. Desai, *J. Phys. Chem. Ref. Data* **15**, 967 (1986).
12. P. D. Desai, *J. Phys. Chem. Ref. Data* **16**, 91 (1987).
13. P. D. Desai, *Int. J. Thermophys.* **8**, 781 (1987).
14. K. Yamaguchi et al., *Mater. Trans. JIM* **35**, 118 (1994).
15. R. Brand et al., *J. Chem. Phys.* **116**, 10386 (2002).
16. J. Font et al., *Mater. Res. Bull.* **30**, 839 (1995).
17. http://www.webelements.com/.
18. D. Bansal et al., *Phys. Rev. B* **94**, 054307 (2016).
19. R. E. Hanneman et al., *J. Appl. Phys.* **36**, 1794 (1965).
20. L. Mañosa et al., *Nat. Mater.* **9**, 478 (2010).
21. E. Stern-Taulats et al., *Phys. Rev. B* **89**, 214105 (2014).
22. D. Matsunami et al., *Nat. Mater.* **14**, 73 (2015).
23. P. Lloveras et al., *Nat. Commun.* **6**, 8801 (2015).
24. E. Stern-Taulats et al., *APL Mater.* **4**, 091102 (2016).
25. J. M. Bermúdez-García et al., *Nat. Commun.* **8**, 15715 (2017).
26. A. Aznar et al., *Nat. Commun.* **8**, 1851 (2017).
27. D. Chandra et al., *Powder Diffraction* **8**, 109 (1993).
28. M. Bée, Quasielastic Neutron Scattering Principles And Applications In Solid State Chemistry, Biology And Materials Science IOP Publishing Ltd (1988).
29. G. B. Guthrie et al., *J. Phys. Chem. Solids* **18**, 53 (1961).
30. J. Font et al., *J. Mater. Res.* **12**, 3254 (1997).
31. F. E. Karasz et al., *J. Phys. Chem. Solids* **20**, 294 (1961).

**Acknowledgments:** We acknowledge the award of beam time from J-PARC (proposal no. 2018U1401) and ANSTO. **Funding:** B.L. W.J.R., and Z.D.Z. were supported by the Hundred Talents Project of CAS. **Author contributions:** B.L. and Z.D.Z. conceived the idea of the work. Y.K., T.K. and K.N. performed the neutron scattering experiments at AMATERAS. S.Y. collected the elastic data at SIKA. D.-H.Y. conducted the inelastic neutron scattering measurements at Pelican. H. W. conducted the DFT calculations. W.J.R. tested the sample. B.L. analyzed all data and wrote the manuscript with discussion and input from all coauthors. **Competing interests:** The authors declare no competing interests. **Data and materials availability:** All data are presented in the main text and supplementary materials.

**Fig. 1. Plastic crystals with CBCE materials for next-generation solid-state refrigeration technology.** (**a**) Fusion enthalpy and solid-state transition enthalpy for typical regular solids as well as plastic crystals: neopentylglycol (NPG), pentaglycerin (PG), pentaerythritol (PE), 2-Amino-2-methyl-1,3-propanediol (AMP), tris (hydroxymethyl) aminomethane (TRIS), 2-Methyl-2-nitro-l-propanol (MNP), 2-Nitro-2-methyl-1,3-propanediol (NMP). (**b**) Compressibility for typical regular solids and plastic crystals. (**c**) $|\Delta S_{max}|$ for leading caloric-effect materials. The values for plastic crystals are evaluated using the Clausius–Clapeyron relation (Table S1).

**Fig. 2. Phase transition of NPG plastic crystal.** (**a**) The molecular configuration and its motion in the ordered and disordered phases. (**b**) The elastic intensity at different temperatures obtained at AMATERAS. (**c**) Incoherent elastic intensity at 2.1 Å$^{-1}$ as a function of temperature recorded with cooling down at SIKA. (**d**) The activation energy ($E_a$) of the isotropic reorientation mode obtained by fitting the temperature dependence of experimental linewidth ($\Gamma$).

**Fig. 3. Reorientational dynamics of molecules.** (**a**) Contour plots of $S(Q,E)$ at five temperatures around $T_t$ obtained with $E_i$ = 2.6363 meV at AMATERAS. (**b**) and (**c**), Incoherent scattering profiles as a function of $E$ at $0.9 \leq Q \leq 1.0$ Å$^{-1}$ at 300 and 320 K, respectively. The multiple-components fitting are also shown. One Lorentzian function at 300 K is needed. However, two at 320 K are involved, where Lorentzian 1 is wider (faster motion) while Lorentzian 2 is narrower (slower motion). (**d**) and (**e**), EISF at 300 and 320 K, respectively determined by repeating the fitting in different $Q$ positions. The EISF is fitted to partial $C_3$ model and isotropic reorientation model as shown in Fig. 2a. (**f**) and (**g**), line width at 300 and 320 K.

**Fig. 4. Anharmonic lattice dynamics of NPG.** (**a**) The INS spectrum at 5 K of NPG obtained with $E_i$ = 23.76 meV at AMATERAS. (**b**) The sliced $S(Q,E)$ for $0.9 \leq Q \leq 1.0$ Å$^{-1}$ at 5, 300 and 320 K for AMATERAS data. The ticks underneath the spectra represent the calculated vibrational frequencies in DFT. The inset shows the temperature dependence of energy of 12.7 meV mode (arrowed) combined the AMATERAS and Pelican data.

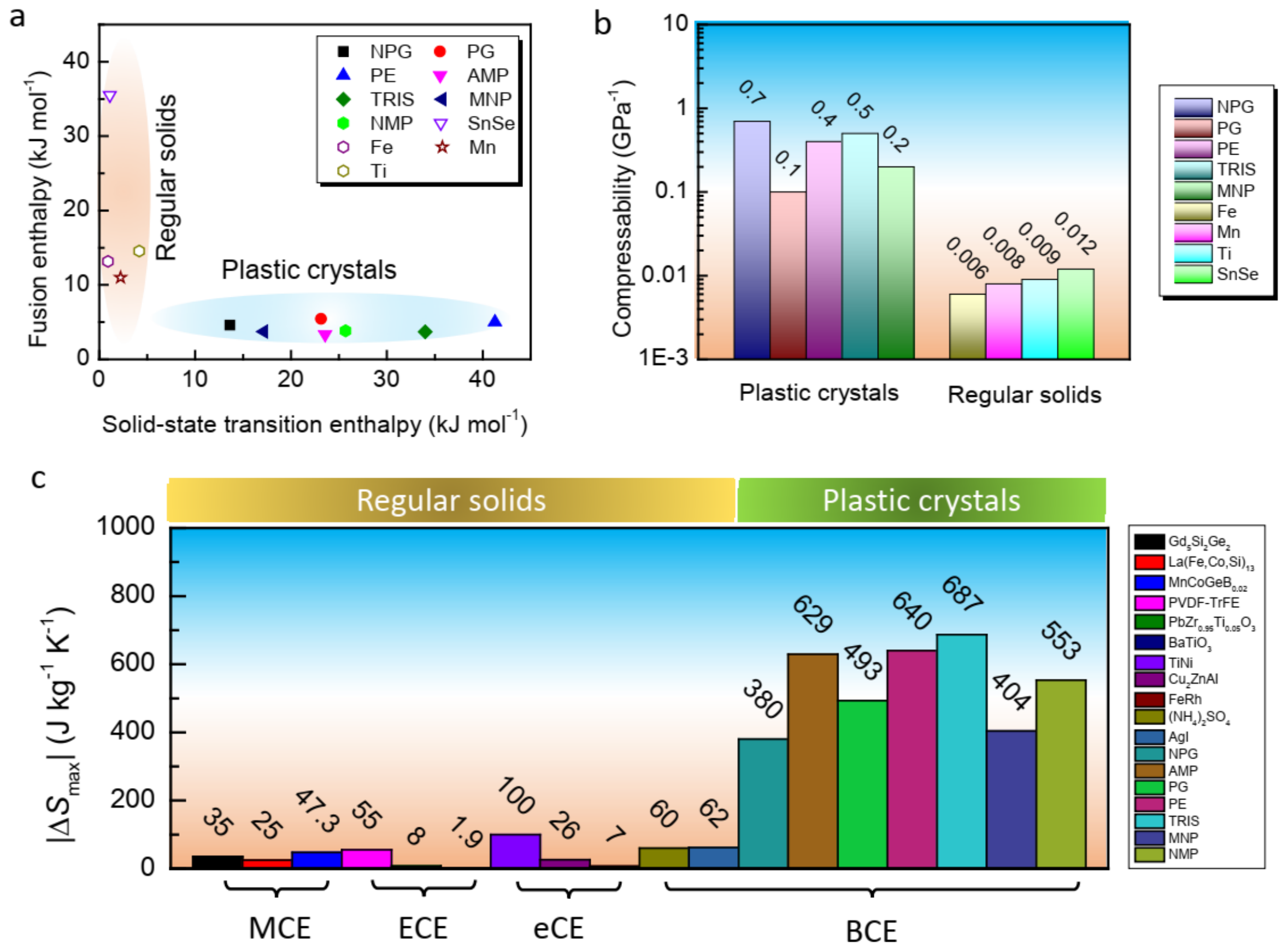

# Figure 1

# Li et al.

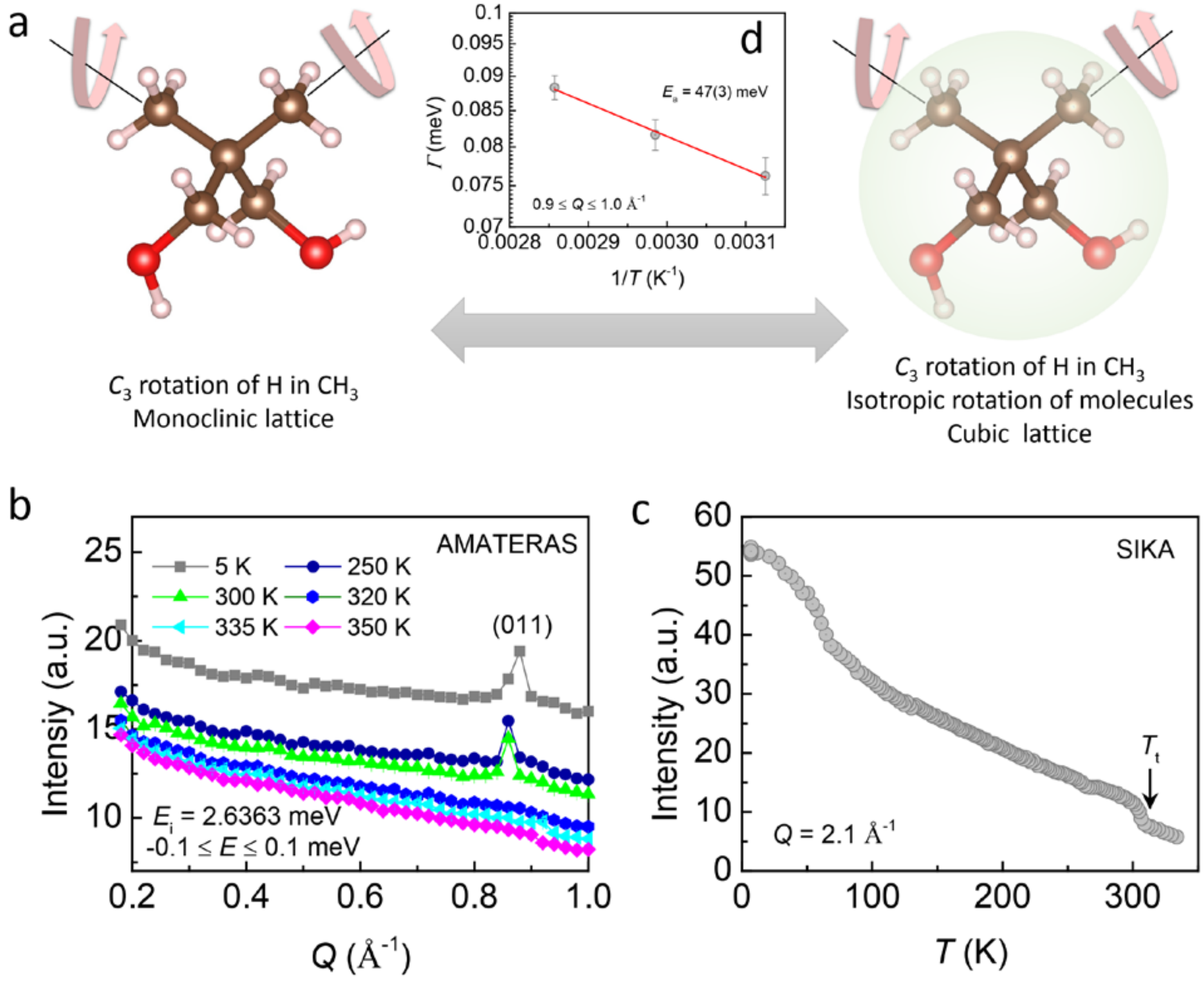

**Figure 2**

**Li et al.**

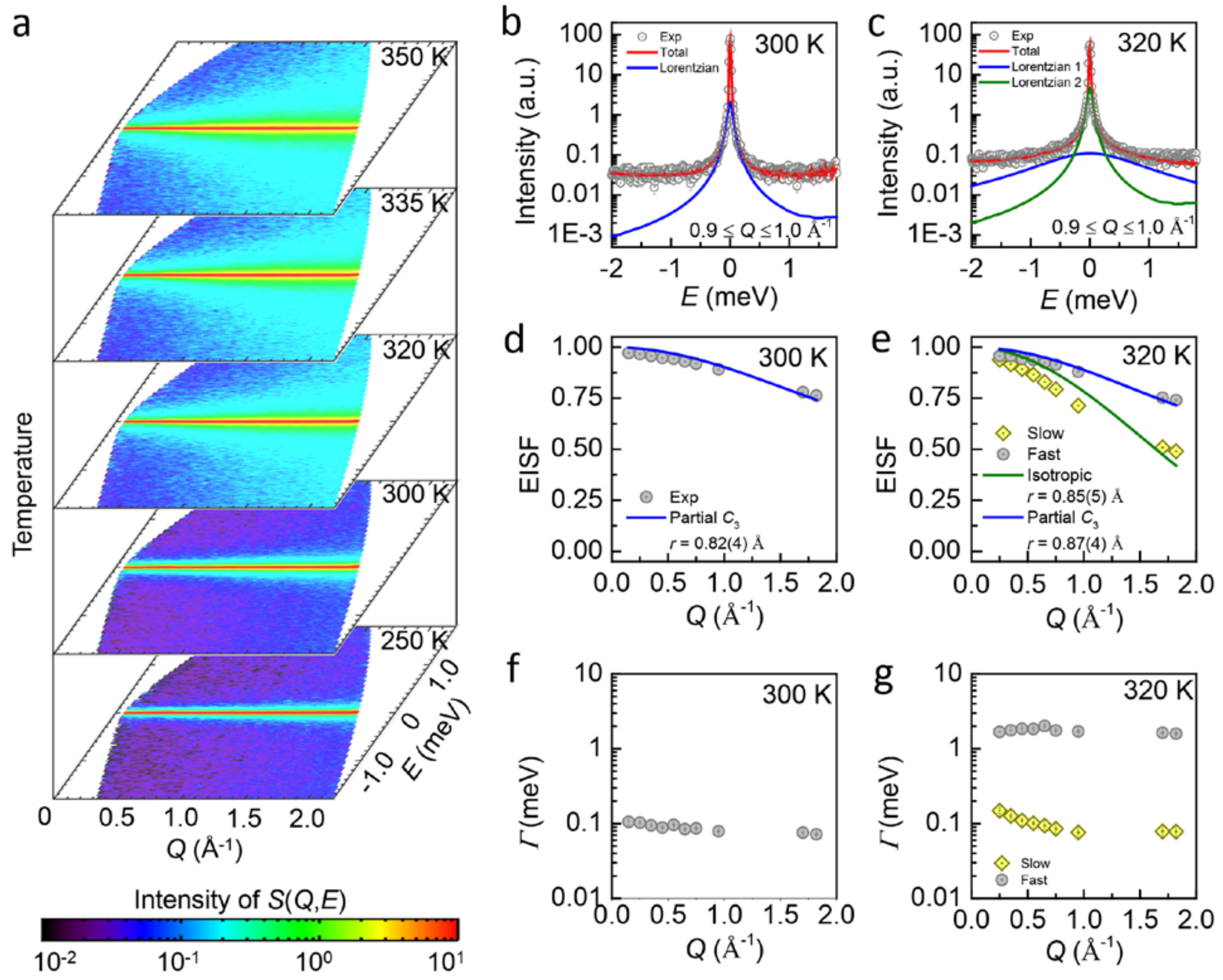

# Figure 3

**Li et al.**

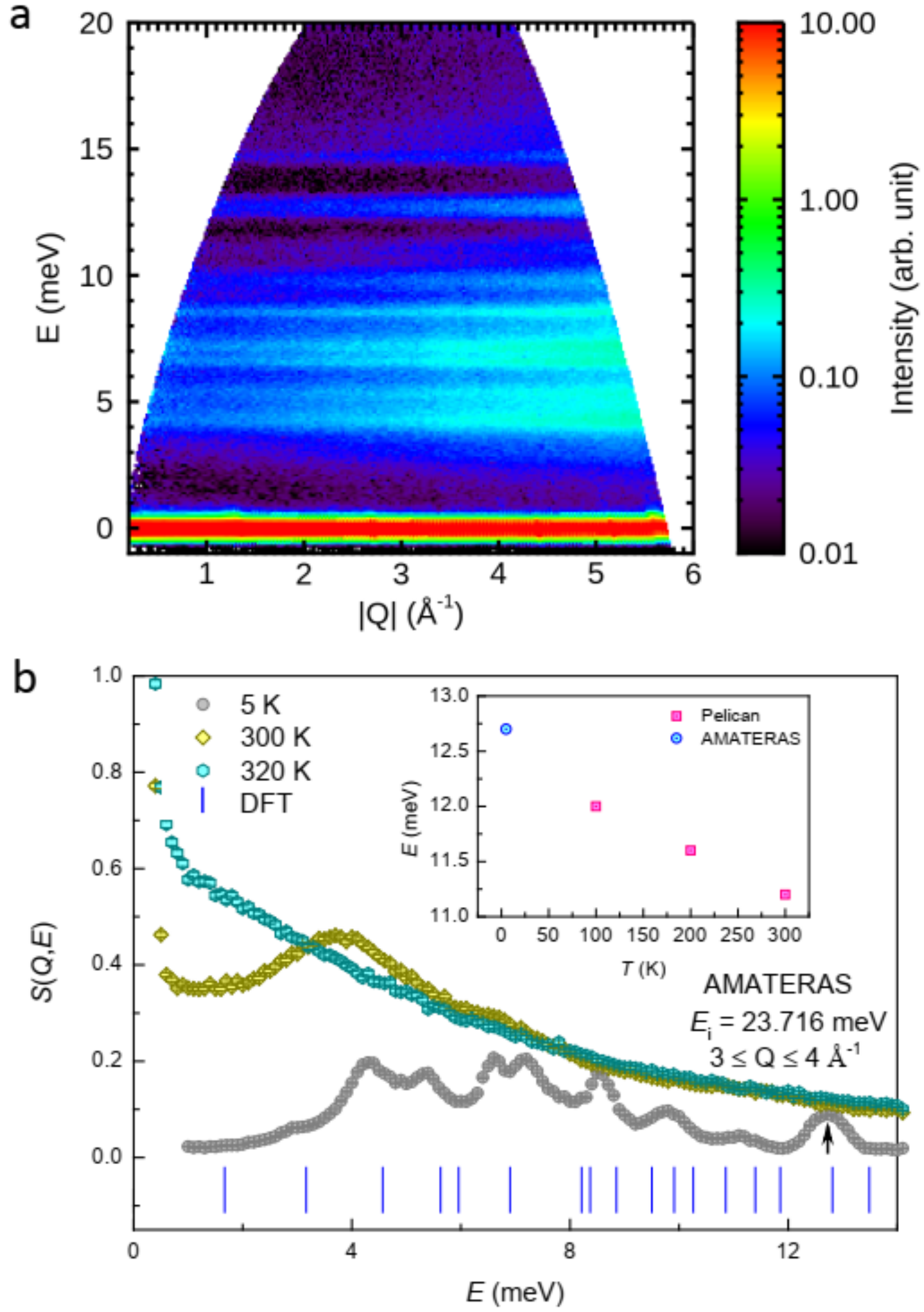

**Figure 4**

**Li et al.**

**Supplementary Materials:**

Materials and Methods

Figures S1-S7

Tables S1-S2

References (*1-48*)

# Supplementary Materials

**Including:**

1. **Materials and Method**
2. **Supplementary figures**
3. **Supplementary tables**
4. **References**

## 1. Materials and Methods

### A) Sample

The powder sample of Neopentylglycol (NPG), whose IUPAC name is 2,2-Dimethyl-1,3-propanediol, was ordered from Sigma-Aldrich. The purity is 99%. The as-ordered powder was checked using lab X-ray diffraction, which suggests it crystallizes in the low-temperature monoclinic phase, as shown in Fig. S1.

### B) Elastic neutron scattering at SIKA

The neutron scattering was performed with the cold-neutron triple-axis spectrometer SIKA at Australian Center for Neutron Scattering (ACNS) of Australian Nuclear Science and Technology (ANSTO) in Australia (*32*). About 6.6 grams of NPG powder was sealed into an aluminium can under helium gas. Incident energy of neutron ($E_i$) and final energy of neutron ($E_f$) were fixed at 5.0 meV to achieve elastic scattering intensities. The monochromater was vertically focused. Collimations were set to be open-open 60\`-60\`. A cooled Be-filer was placed at scattered side to cut off the higher order contamination, such as lambda over two and lambda over three. Constant $Q$ scan were made at 2.1 Å$^{-1}$ to collect incoherent elastic neutron scattering intensities.

### C) Inelastic neutron scattering at AMATERAS

Multi-$E_i$ time-of-flight inelastic neutron scattering measurements were performed at the cold-neutron disk chopper spectrometer BL14 AMATERAS of J-PARC in Japan (33,34). The powder sample around 0.29 gram was wrapped with aluminum foil and further sealed into an aluminum can with indium wire. Such a geometry assures about 80% transmission of neutrons, which is crucial to obtain high-quality quasi-elastic neutron scattering data. A cryostat was used to access the broad temperature region. The choppers configurations were set to select $E_i$ of 23.716, 5.9313 and 2.6363 meV. The corresponding energy resolution are 0.42, 0.078, and 0.026 meV, respectively, as shown in Fig. S2. The data reduction was completed by using Utsusemi suite (35). The resulted $S(Q,E)$ data was visualized in Mslice of DAVE (36). The contour plots of $S(Q,E)$ with $E_i$ of 23.716 and 5.9313 meV are shown in Fig. S3.

### D) Inelastic neutron scattering at PELICAN

The inelastic neutron scattering experiment was performed using the cold-neutron time-of-flight spectrometer PELICAN at the ACNS of ANSTO in Australia (37,38). The instrument was configured for 5.96 Å incident neutrons, affording an energy resolution of 0.065 meV at the elastic line for this experiment. The powder sample NPG was loaded into a flat-plate aluminium can with 0.2 mm sample thickness, under dry $N_2$ gas environment. The sample was oriented 135° to the incident neutron beam. A top-loading cryostat was used to maintain the sample temperature from 10 K to 350 K. A background spectrum for an empty can and instrument resolution function from a standard vanadium sample under the same configuration as the measurements for sample were also collected. The spectrum for vanadium standard was also used for detector normalization. All data reduction and manipulations including background subtraction and detector normalization were done using the Large Array Manipulation Program (LAMP) (39). The general density of state (GDOS) is shown in Fig. S4.

### E) QENS analysis and reorientation dynamics

With $E_i$ of 2.6363 meV at AMATERAS, the supreme energy resolution allows us to thoroughly examine the quasi-elastic scattering information. The $S(Q,E)$ data were converted into one-dimensional data at specific $Q$ points. $Q$ slicing was done at [0.1,0.2], [0.2,0.3], [0.3,0.4], [0.4,0.5], [0.5,0.6], [0.6,0.7], [0.7,0.8] [0.9,1.0], [1.65,1.75] and

[1.78,1.88] Å$^{-1}$, respectively. These spectra were fitted in PAN of DAVE by including a Lorentzian function below $T_t$ while two above $T_t$, a delta function, and a constant background, which are convoluted to the instrumental resolution. They describe the quasi-elastic scattering, incoherent elastic scattering, and background, respectively. The 5 K data were used as resolution functions for individual $Q$ position.

At 300 K, there is only one Lorentzian function needed, whose elastic incoherent scattering factor (EISF) is defined by the ratio of intensity as

$$\text{EISF (300 K)} = \frac{I_{Elastic}}{I_{Elastic} + I_{QENS}}$$

Since there are two Lorentzian functions found at 320 K, the definitions are modified to be

$$\text{EISF (320 K, slow)} = \frac{I_{Elastic}}{I_{Elastic} + I_{QENS,slow}}$$

$$\text{EISF (320 K, fast)} = \frac{I_{Elastic} + I_{QENS,slow}}{I_{Elastic} + I_{QENS,slow} + I_{QENS,fast}}$$

The EISF at 300 K is well reproduced by the partial three fold ($C_3$) reorientation mode of hydrogen atoms in methyl group, given as

$$\text{EISF (partial } C_3) = \frac{1}{2} + \frac{1}{2} * \frac{1}{3}\left[1 + 2 * \sin(\sqrt{3} * Q * r)/(\sqrt{3} * Q * r)\right]$$

Here, $r$ is the radius of the circle on which the hydrogen atoms rotate. The fitted value is 0.82(4) Å, which is closed to the crystallographic data, as shown in Fig. S5. At 320 K, the EISF of the fast mode can be also fitted with this model. The resulted value is 0.87(4) Å. While the slow mode can be reproduced using the isotropic rotation model, whose EISF given as

$$\text{EISF (isotropic)} = [\sin(Q * r)/(Q * r)]^2$$

The resulted r is 0.85(5) Å, which approximates the configuration where the gravity of incoherent scattering objects are on carbon atoms lying the corners of the tetrahedron, as shown in Fig. S6.

The interplay between the reorientational modes and phonons is described in Fig. S7, where it can be seen that the phonons become over-broadened as the isotropic reorientational mode of the molecules are is activated above $T_t$.

**F) DFT calculations**

The density functional theory calculations (DFT) were performed with the Vienna Ab-initio Simulation Package (VASP) (40,41). The spin-polarized generalized gradient approximation (GGA) (42) was used for the description of the exchange-correlation interaction among electrons. We treated C-2*s*2*p*, H-1*s* and O-2*s*2*p* as valence states and adopted the projector-augmented wave (PAW) pseudopotentials to represent the effect of their ionic cores (43,44). Spin-orbit coupling (SOC) was not included in the calculation given that SOC of C, H and O are very weak. We sampled the Brillouin zone by adopting the Γ-centered Monkhorst-Pack (45) method with a density of about 2π×0.03 Å-1 in all calculations. Brillouin zone integrations are performed with a Gaussian broadening of 0.05 eV during all calculations (46). For the correct description of weak interaction between each molecules, the non-local van der Waals correction is included using the DFT-D3 method (47). Finite differences are used to obtain the forces by displacing each ion in three Cartesian directions, from which the Hessian matrix and vibrational frequencies are determined. The energy cutoff for the plane-wave expansion was 600 eV which result in good convergence of the computed ground-state properties. Structures were optimized with a criterion that the atomic force on each atom becomes weaker than 0.01 eV/Å and the energy convergence is better than $10^{-6}$ eV.

## 2. Figure S1-S7

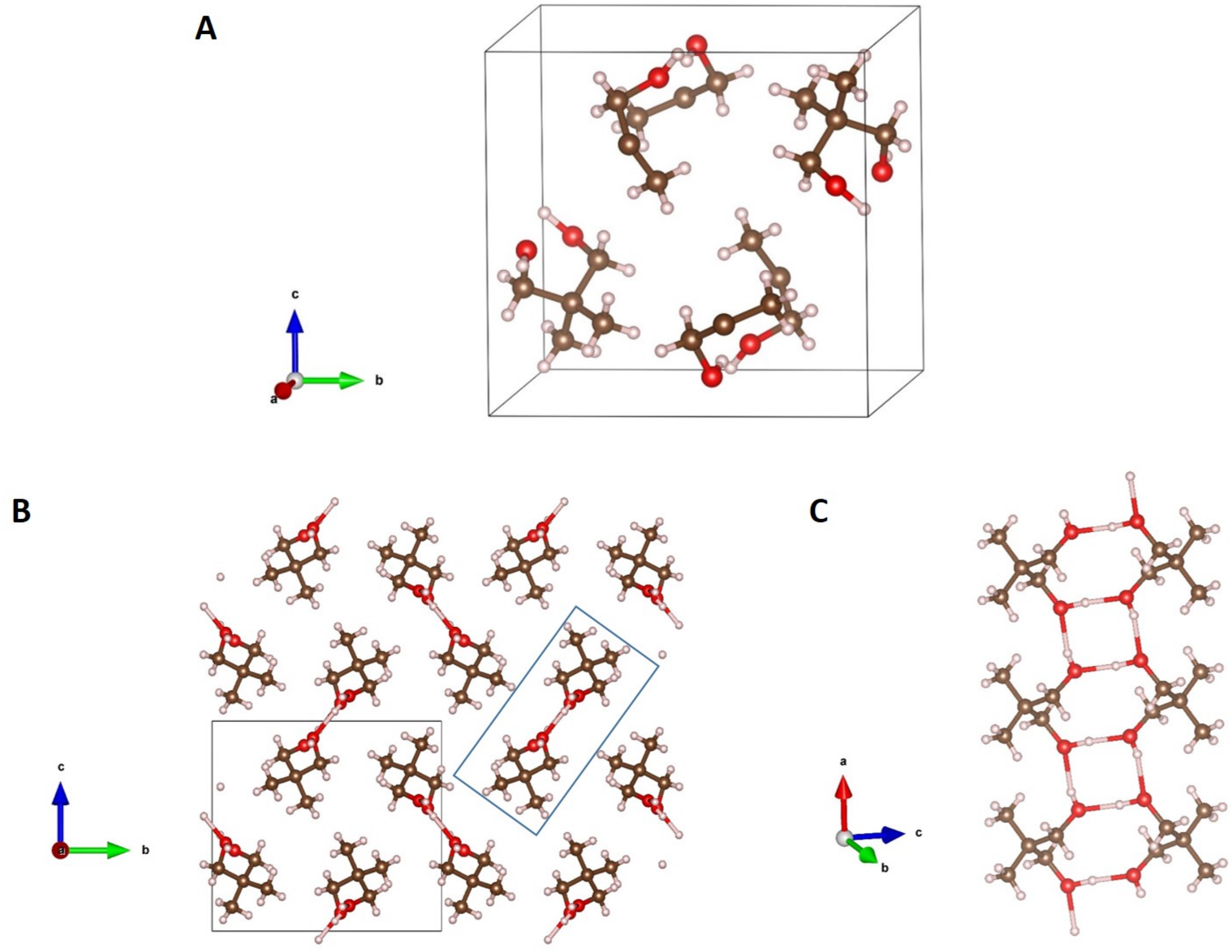

**Figure S1. The crystal structure of the monoclinic phase.** (**A**) The unit cell. (**B**) The extended view showing hydrogen bonds. The rectangle highlights the hydrogen-bond ladders. (**C**) The hydrogen-bond ladder repeating along the *a* axis.

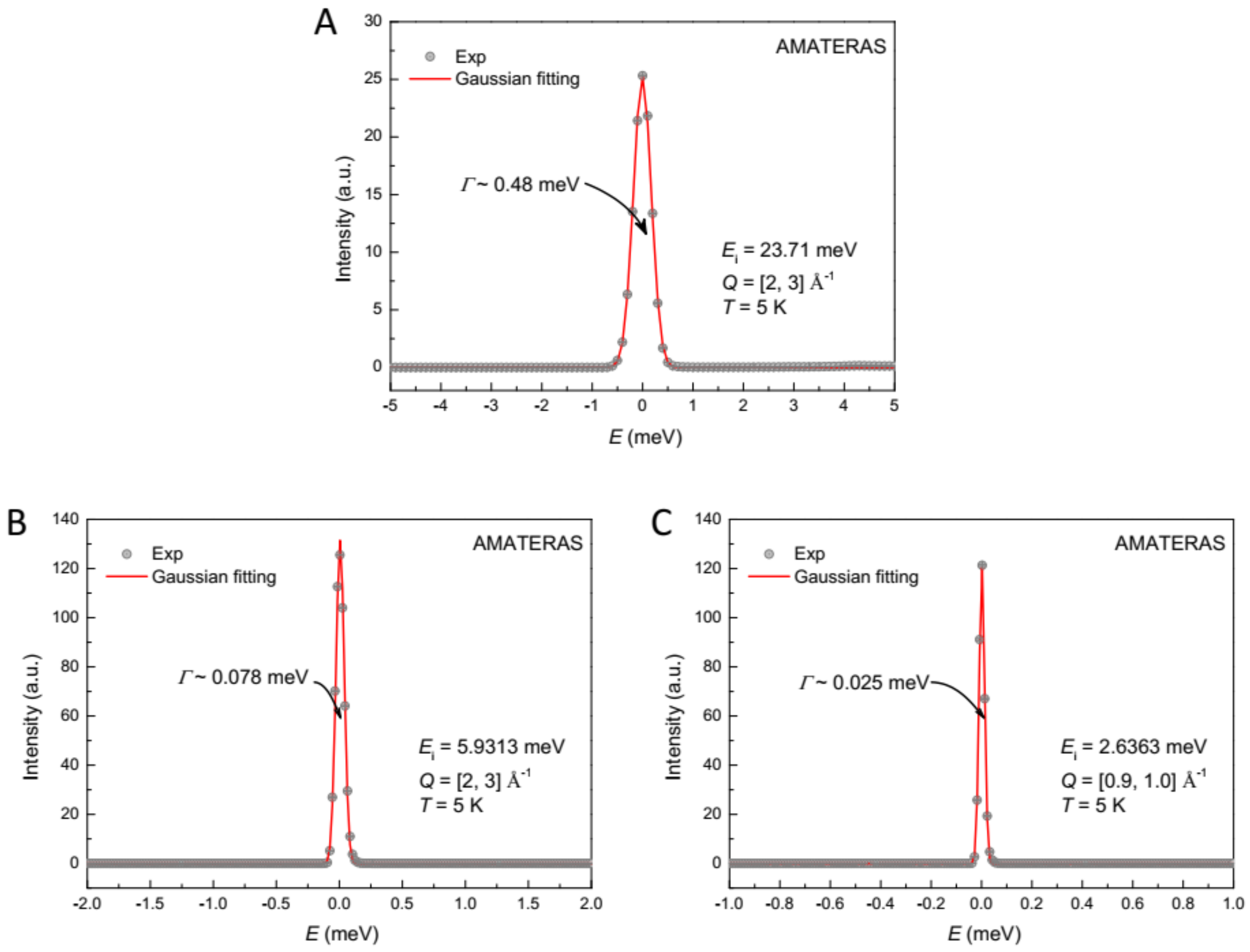

**Figure S2. Energy resolutions of AMATERAS.** (**A**) At $E_i$ = 23.71 meV, the resolution is 0.48 meV. (**B**) At $E_i$ = 5.9313 meV, the resolution is 0.078 meV. (**C**) At $E_i$ = 2.6363 meV, the resolution is 0.025 meV.

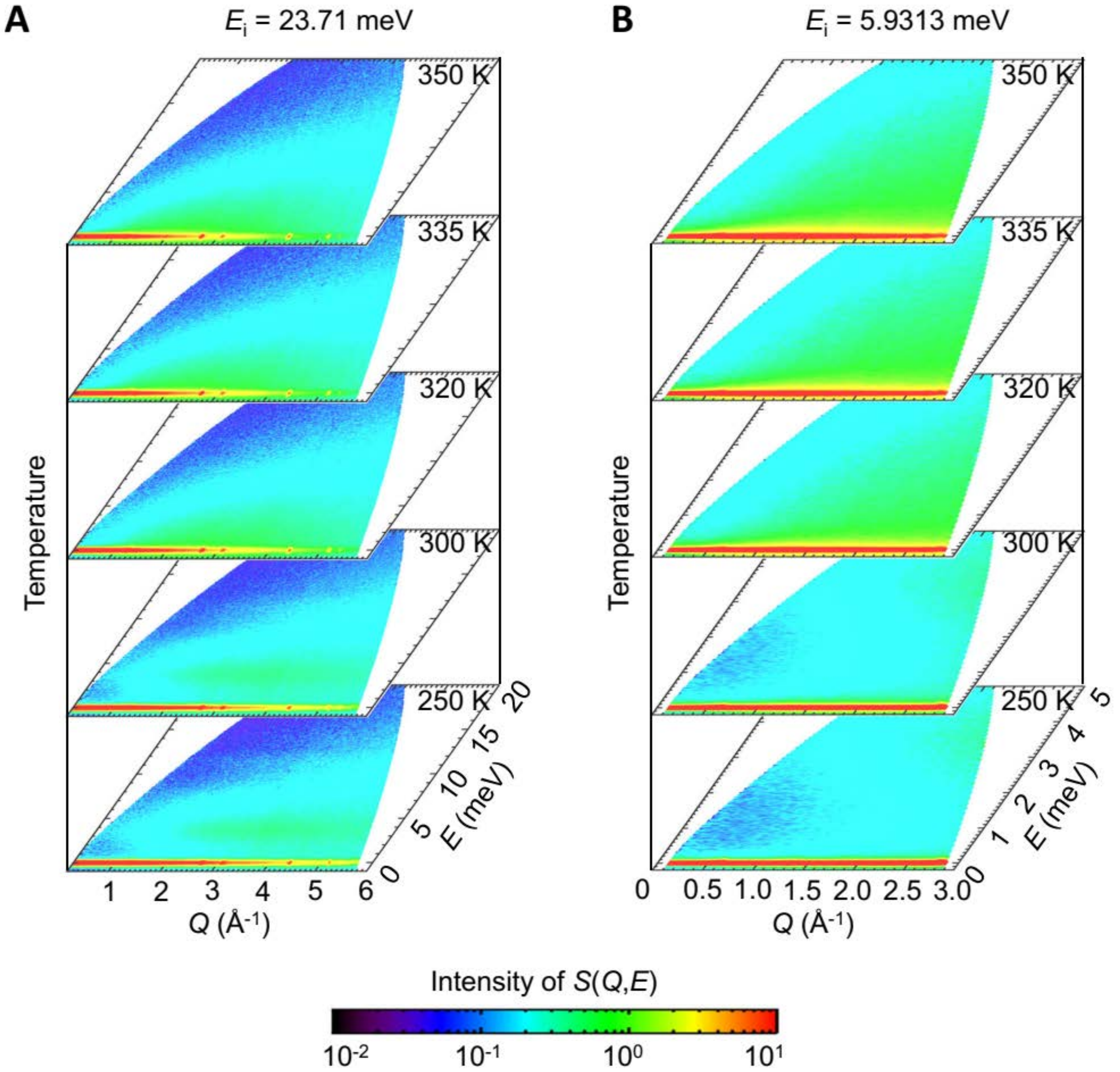

**Figure S3. *S*(*Q*,*E*) contour plots obtained at AMATERAS.** (**A**) $E_i$ = 23.71 meV. (**B**) $E_i$ = 5.9313meV.

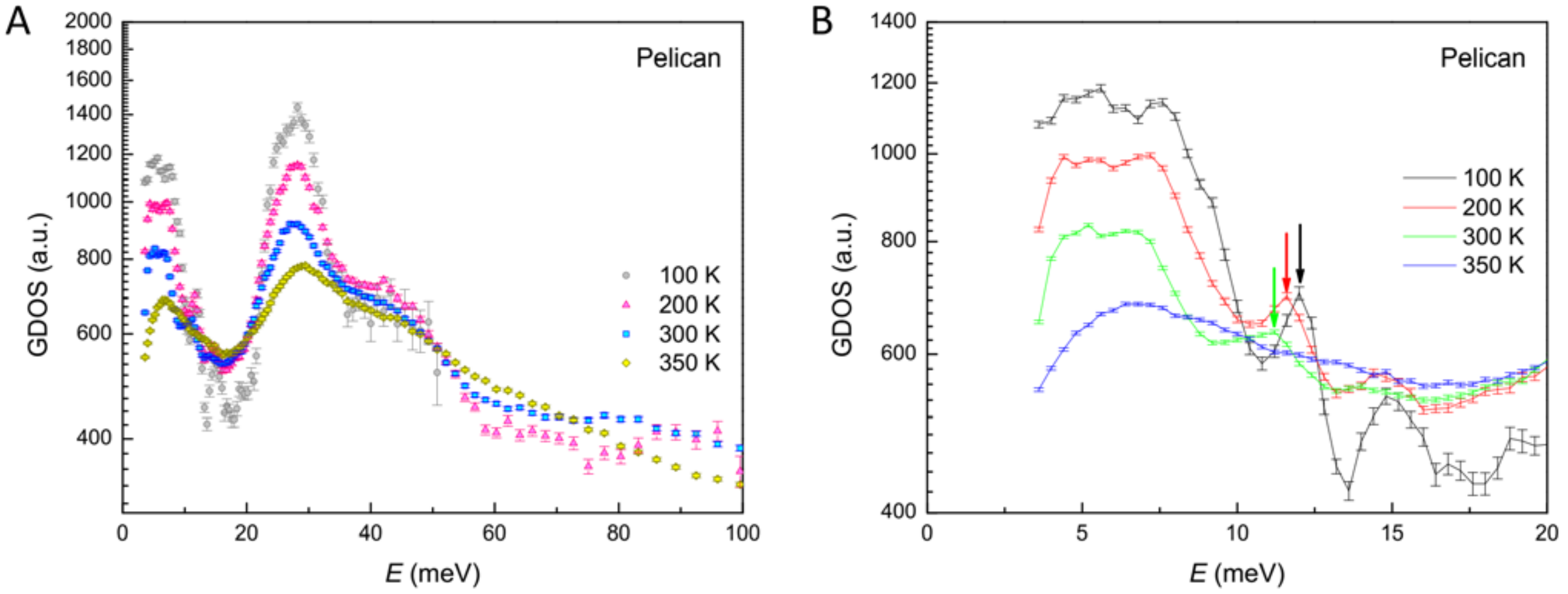

**Figure S4. General density of state (GDOS) obtained at Pelican.** (**A**) The GDOS up to 100 meV at 100, 200, 300 and 350 K. (**B**) The highlight of the lower energy region. The arrow point out the peak positions of the mode around 12.7 meV.

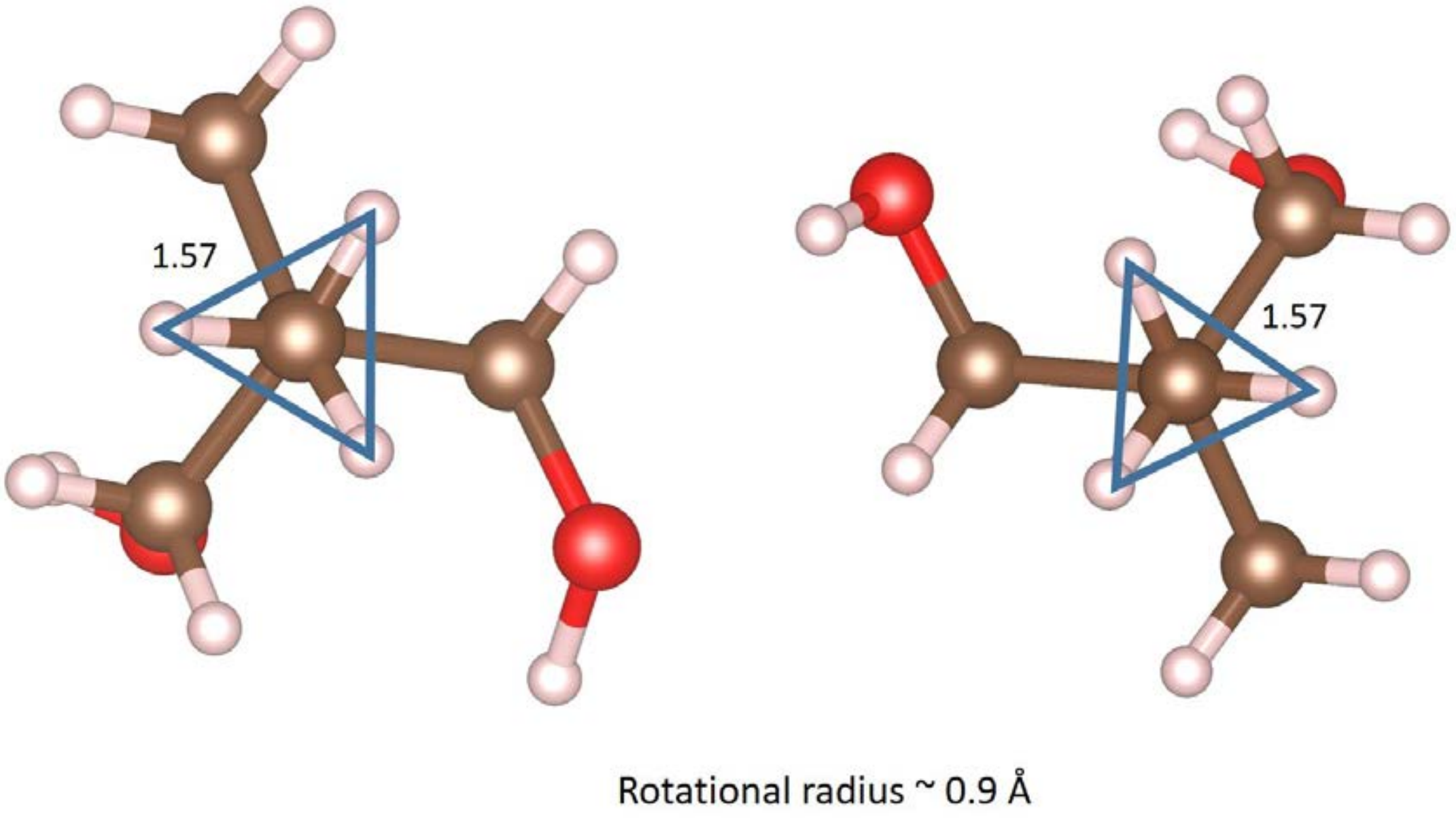

**Figure S5. The geometry of methyl group in NPG.**

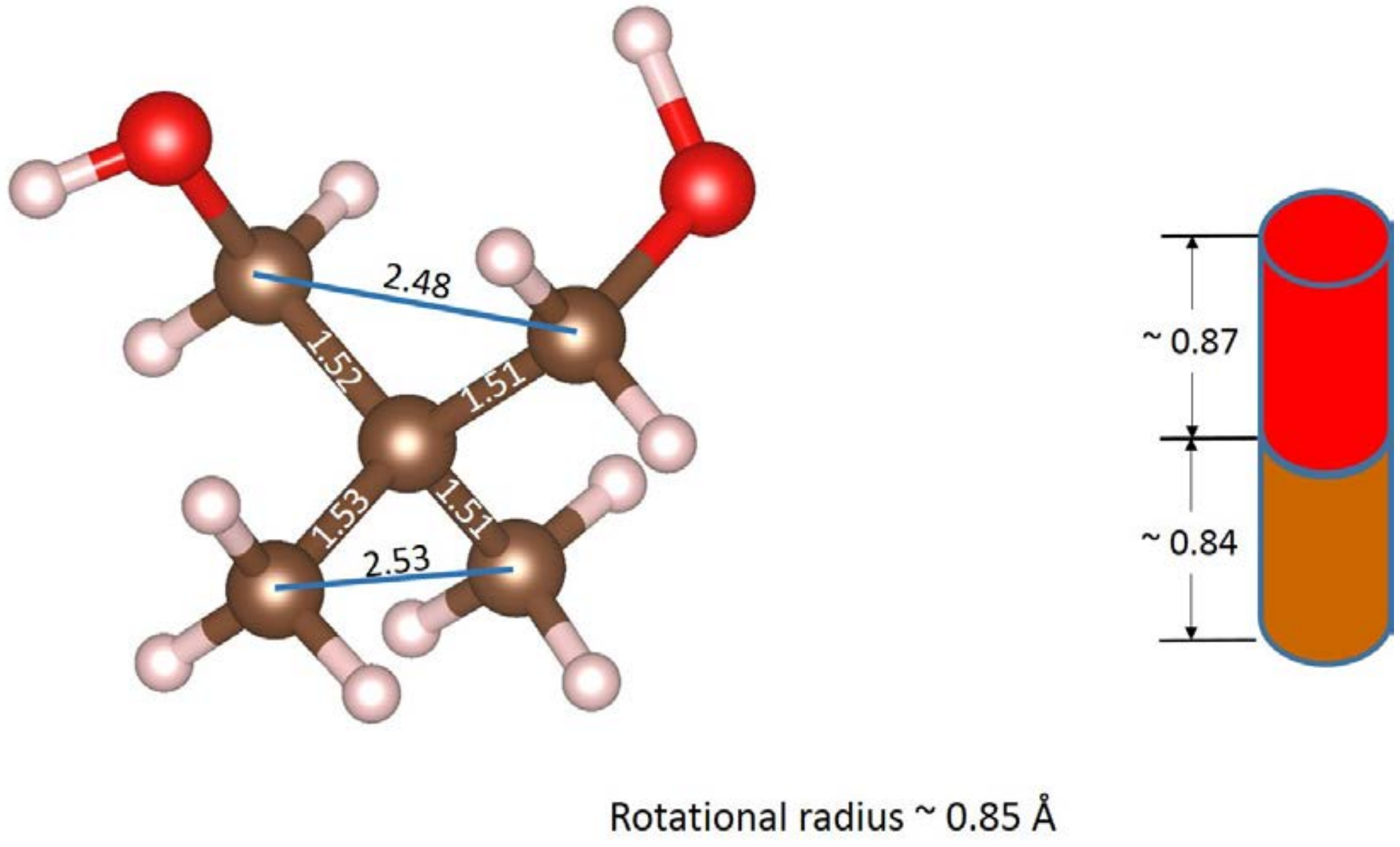

**Figure S6. The geometry of NPG molecule mimicked by the pole shape.**

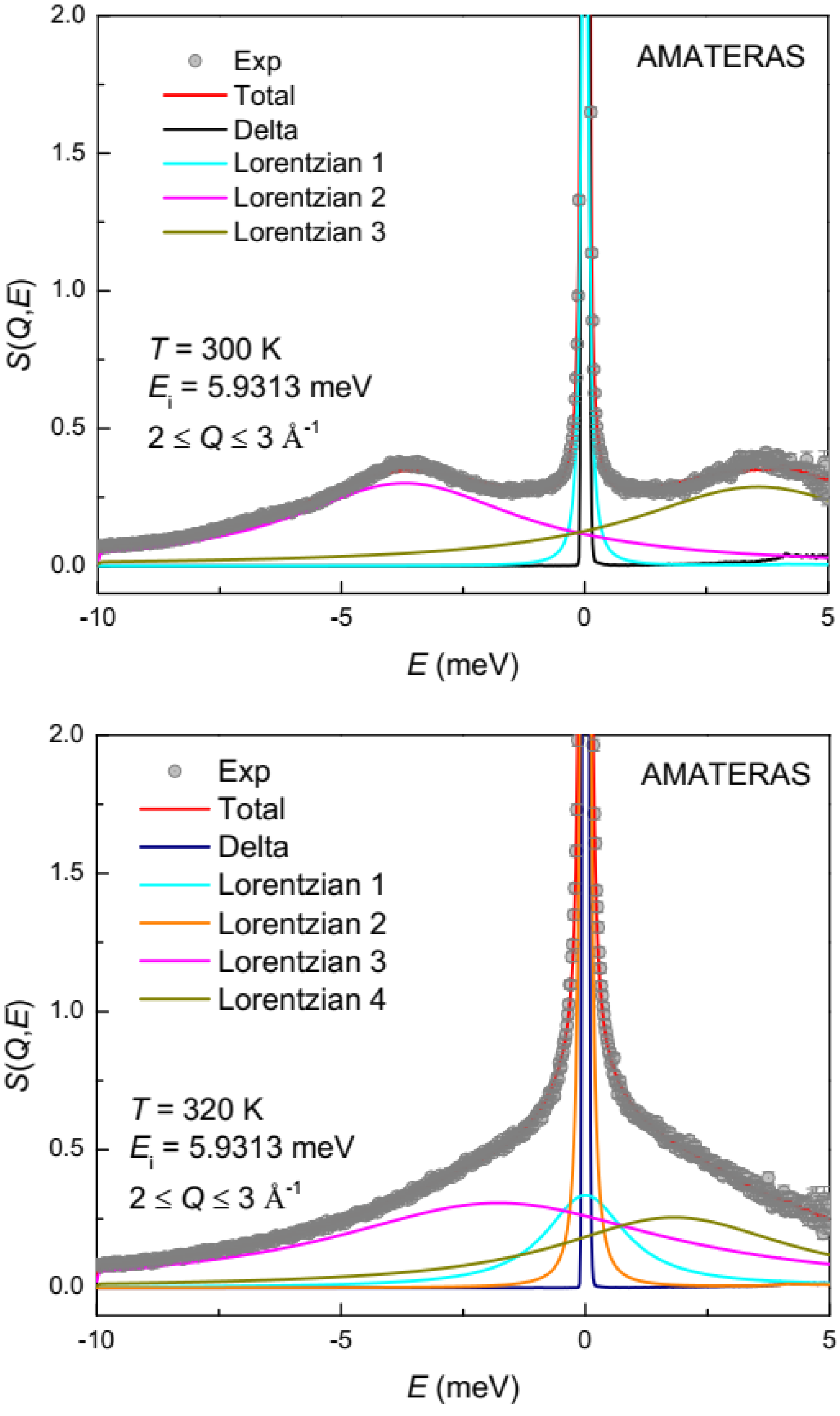

**Figure S7. Multi-component fitting of $S(Q,E)$ data at $2 \leq Q \leq 3$ Å$^{-1}$ with $E_i$ = 5.9313 meV.** (**A**) 300 K. (**B**) 320 K.

## 3. Table S1 to S2

**Table S1. Evaluation of entropy changes using the Clausius–Clapeyron relation on a few plastic crystals. (16)**

| Plastic crystals | $M$ (g/mol) | $T_t$ (K) | $dP/dT_t$ (MPa/K) | $\Delta V$ ($cm^3$/mol) | $\Delta V$ ($m^3$ $kg^{-1}$) | $\Delta S$ (J $kg^{-1}$ $K^{-1}$) |
|---|---|---|---|---|---|---|
| NPG | 104.15 | 313 | 7.5 | 5.4 | 5.18483E-5 | 388.86222 |
| PG | 120.15 | 354 | 9.4 | 6.3 | 5.24345E-5 | 492.8839 |
| PE | 136.15 | 461 | 8.3 | 10.5 | 7.71208E-5 | 640.10283 |
| AMP | 105.14 | 353 | 11.6 | 5.7 | 5.42134E-5 | 628.87578 |
| TRIS | 121.13 | 407 | 14.6 | 5.7 | 4.70569E-5 | 687.03046 |
| MNP | 119.12 | 308 | 10.7 | 4.5 | 3.7777E-5 | 404.21424 |
| NMP | 135.12 | 352 | 7.4 | 10.1 | 7.47484E-5 | 553.13795 |

**Table S2. List of some plastic crystals (16-26,48).**

| No. | Formula | Name | Transition temperature （K） | Entropy （$Jmol^{-1}K^{-1}$） |
|---|---|---|---|---|
| 1 | $CH_4$ | Methane | 20.5 | 3.18 |
| 2 | $CF_4$ | Tetrafluoromethane | 76.2 | 19.4 |
| 3 | $CCl_4$ | Carbon tetrachloride | 225.5 | 20.33 |
| 4 | $CBr_4$ | Tetrabromomethane | 320 | 20.82 |
| 5 | $C(CH_3)_4$ | 2,2-Dimethylpropane | 140 | 18.4 |
| 6 | $C(CH_3)_3Cl$ | cholorobutane | 219 | 36.38 |
| 7 | $C(CH_3)Cl_3$ | 1,1,1-trichloro-2-methy lpropane | 224 | 32.91 |
| 8 | $C(CH_3)_3C_2H_5$ | 3-methyl pentane | 126.8 | 42.53 |
| 9 | $C(CH_3)_3SH$ | trimethanethiol | 151.6 | 26.81 |
| 10 | $C_2Cl_6$ | hexachloroethane | 344.6 | 23.84 |
| 11 | $Si_2(CH_3)_6$ | hexamethyldisilane | 221.9 | 43.91 |
| 12 | $(CH_3)_2CHCH(CH_3)_2$ | 2,3-Dimethylbutane | 136.1 | 47.72 |
| 13 | $C_4H_8$ | Cyclobutane | 145.7 | 39.14 |
| 14 | $C_5H_{10}$ | Cyclopentane | 122.4 | 39.81 |
| 15 | $C_6H_{12}$ | Cyclohexane | 186 | 35.92 |
| 16 | $HOCH(CH_2)_5$ | Cyclohexanol | 263 | 35.16 |
| 17 | $C_{10}H_{16}O$ | 1,7,7-trimethylbicyclo | 250 | 31.78 |
| 18 | $(CH_3)_2C(CH_2OH)_2$ | 2,2-Dimethyl-1,3-propa nediol (NPG) | 313 | 43.24 |
| 19 | $CH_3C(CH_2OH)_3$ | 2-Hydroxymethyl-2-me thyl-1,3-propanediol (PG) | 354 | 65.23 |
| 20 | $C(CH_2OH)_4$ | 2,2-Bis(hydroxymethyl) 1,3-propanediol (PE) | 461 | 90.28 |
| 21 | $(CH_3)C(NH_2)(CH_2OH)_2$ | 2-Amino-2-methyl-1,3-propanediol (AMP) | 353 | 67.29 |
| 22 | $(NH_2)C(CH_2O$ | 2-Amino-2-hydroxymet | 404 | 84.89 |
| | H)$_3$ | hyl-1,3-propanediol | | |
| 23 | $(CH_3)_2C(NO_2)(CH_2OH)$ | 2-Methyl-2-nitro-1-propanol | 308 | 55.62 |
| 24 | $(CH_3)C(NO_2)(CH_2OH)_2$ | 2-Nitro-2-methyl-1,3-propanediol | 352 | 73.19 |
| 25 | $(CH_3)_3CCH_2OH$ | 2,2-dimethyl-1-propanol | 242 | 18.4 |
| 26 | $HC(CH_2OH)_3$ | 2-Hydroxymethyl-1,3-propanediol | 354 | 65.24 |
| 27 | $(CH_3)_3CCOOH$ | 2,2-Dimethylpropanoic acid | 280 | 36.15 |
| 28 | $(CH_2OH)_3CCOOH$ | tris(hydroxymethyl)ethanoic acid | 397 | 77.37 |
| 29 | $C_5H_{11}ClO_4$ | monochloro pentaerythritol | 334 | 69 |
| 30 | $C_5H_{11}FO_4$ | monofluoro pentaerythritol | 341 | 56.46 |
| 31 | $C_5H_{10}F_2O_2$ | difluoro pentaerythritol | 293 | 43.07 |
| 32 | $C_5H_{11}F_4$ | tetrafluoro pentaerythritol | 247 | 53.02 |
| 33 | $C_5H_{13}NO_3$ | monoamine pentaerythritol | 359 | 69.42 |
| 34 | $C_5H_{14}N_2O_2$ | diamine pentaerythritol | 341 | 77.37 |
| 35 | $SF_6$ | Sulfur hexafluoride | 94.3 | 17.02 |
| 36 | $PtF_6$ | Platinum hexafluoride | 276 | 32.2 |
| 37 | NaSn | Sodium tin | 757 | 4.95 |
| 38 | CsPb | Cesium lead | 869 | 7.15 |
| 39 | HCl | hydrogen chloride | 98.4 | 12.09 |
| 40 | $H_2S$ | Hydrogen sulfide | 103.5 | 14.85 |
| 41 | $N_2$ | Nitrogen gas | 35.6 | 6.44 |
| 42 | $SiH_4$ | Silane | 63.45 | 10.91 |
| 43 | $GeH_4$ | Germane | 73.2 | 7.44 |
| 44 | $C_{60}$ | fullerene | 260 | 30 |

## 4. References

1. The Importance of Energy Efficiency in the Refrigeration, Air-conditioning and Heat Pump Sectors. United Nations Environmental Programme, May 2018. http://conf.montreal-protocol.org/meeting/workshops/energy-efficiency/presession/breifingnotes/briefingnote-a_importance-of-energy-efficiency-in-the-refrigeration-air-conditioning-and-heat-pump-sectors.pdf

2. Savings and benefits of global regulations for energy efficient products, Ecofys, 2015.

3. X. Moya, S. Kar-Narayan and N. D. Mathur, Caloric materials near ferroic phase transitions, Nat. Mater. 13, 439 (2014). Doi: https://doi.org/10.1038/nmat3951.

4. V. Franco, J. S. Blázquez, J. J. Ipus, J. Y. Law, L. M. Moreno-Ramírez, A. Conde, Magnetocaloric effect: From materials research to refrigeration devices, Prog. Mater. Sci. 93, 112 (2018). Doi: https://doi.org/10.1016/j.pmatsci.2017.10.005.

5. J. F. Scott, Electrocaloric Materials, Annu. Rev. Mater. Res. 41, 229 (2011). Doi: https://doi.org/10.1146/annurev-matsci-062910-100341.

6. L. Mañosa and A. Planes, Materials with Giant Mechanocaloric Effects: Cooling by Strength, Adv. Mater. 29, 1603607 (2017). Doi: https://doi.org/10.1002/adma.201603607.

7. L. A. K. Staveley, Phase Transitions in Plastic Crystals, Annu. Rev. Phys. Chem. 13, 351 (1962). Doi: https://doi.org/10.1146/annurev.pc.13.100162.002031.

8. D. Chandra, W. Ding, R. A. Lynch, J. J.Tomilinson, Phase transitions in “plastic crystals”, J. Less Common. Met. 168, 159 (1991). Doi: https://doi.org/10.1016/0022-5088(91)90042-3.

9. E. Murrill and L.Breed, Solid—solid phase transitions determined by differential scanning calorimetry: Part I. Tetrahedral substances, Thermochim. Acta 1, 239 (1970). Doi: https://doi.org/10.1016/0040-6031(70)80027-2.

10. J. L. Tamarit, M. A. Pérez-Jubindod and M. R. de la Fuente, Dielectric studies on orientationally disordered phases of neopentylglycol ($(CH_3)_2C(CH_2OH)_2$) and tris(hydroxymethyl aminomethane) ($(NH_2)C(CH_2OH)_3$), J. Phys. Condens. Matter 9, 5469 (1997). Doi: https://doi.org/10.1088/0953-8984/9/25/014.

11. P. D. Desai, Thermodynamic Properties of Iron and Silicon, J. Phys. Chem. Ref. Data 15, 967 (1986). Doi: https://doi.org/10.1063/1.555761.

12. P. D. Desai, Thermodynamic Properties of Manganese and Molybdenum, J. Phys. Chem. Ref. Data 16, 91 (1987). Doi: https://doi.org/10.1063/1.555794.

13. P. D. Desai, Thermodynamic properties of titanium, Int. J. Thermophys. 8, 781 (1987). Doi: https://doi.org/10.1007/BF00500794.

14. K. Yamaguchi, K. Kameda, Y. Takeda and K. Itagaki, Measurements of High Temperature Heat Content of the II-VI and IV-VI (II: Zn, Cd IV: Sn, Pb VI: Se, Te) Compounds, Mater. Trans. JIM 35, 118 (1994). Doi:

15. R. Brand, P. Lunkenheimer, and A. Loidl, Relaxation dynamics in plastic crystals, J. Chem. Phys. 116, 10386 (2002). Doi: https://doi.org/10.1063/1.1477186.

16. J. Font, J. Muntasell and E.Cesari, Plastic crystals: Dilatometric and thermobarometric complementary studies, Mater. Res. Bull. 30, 839 (1995). Doi: https://doi.org/10.1016/0025-5408(95)00055-0.

17. http://www.webelements.com/.

18. D. Bansal, J. Hong, C. W. Li, A. F. May, W. Porter, M. Y. Hu, D. L. Abernathy, and O. Delaire, Phonon anharmonicity and negative thermal expansion in SnSe, Phys. Rev. B 94, 054307 (2016). Doi: https://doi.org/10.1103/PhysRevB.94.054307.

19. R. E. Hanneman and H. C. Gatos, Relationship between Compressibility and Thermal Expansion Coefficients in Cubic Metals and Alloys, J. Appl. Phys. 36, 1794 (1965). Doi: https://doi.org/10.1063/1.1703136.

20. L. Mañosa, D. González-Alonso, A. Planes, E. Bonnot, M. Barrio, J.-L. Tamarit, S. Aksoy and M. Acet, Giant solid-state barocaloric effect in the Ni–Mn–In magnetic shape-memory alloy, Nat. Mater. 9, 478 (2010). Doi: https://doi.org/10.1038/nmat2731.

21. E. Stern-Taulats, A. Planes, P. Lloveras, M. Barrio, J.-L. Tamarit, S. Pramanick, S. Majumdar, C. Frontera, and L. Mañosa, Barocaloric and magnetocaloric effects in Fe49Rh51, Phys. Rev. B 89, 214105 (2014). Doi: https://doi.org/10.1103/PhysRevB.89.214105.

22. D. Matsunami, A. Fujita, K. Takenaka and M. Kano, Giant barocaloric effect enhanced by the frustration of the antiferromagnetic phase in Mn3GaN, Nat. Mater. 14, 73 (2015). Doi: https://doi.org/10.1038/nmat4117.

23. P. Lloveras, E. Stern-Taulats, M. Barrio, J.-L. Tamarit, S. Crossley, W. Li, V. Pomjakushin, A. Planes, L. Mañosa, N. D. Mathur and X. Moya, Giant barocaloric effects at low pressure in ferrielectric ammonium sulphate, Nat. Commun. 6, 8801 (2015). Doi: https://doi.org/10.1038/ncomms9801.

24. E. Stern-Taulats, P. Lloveras, M. Barrio, E. Defay, M. Egilmez, A. Planes, J.-L. Tamarit, L. Mañosa, N. D. Mathur, and X. Moya, Inverse barocaloric effects in ferroelectric BaTiO3 ceramics, APL Mater. 4, 091102 (2016). Doi: https://doi.org/10.1063/1.4961598.

25. J. M. Bermúdez-García, M. Sánchez-Andújar, S. Castro-García, J. López-Beceiro, R. Artiaga and M. A. Señarís-Rodríguez, Giant barocaloric effect in the ferroic organic-inorganic hybrid [TPrA][Mn(dca)3] perovskite under easily accessible pressures, Nat. Commun. 8, 15715 (2017). Doi: https://doi.org/10.1038/ncomms15715.

26. A. Aznar, P. Lloveras, M. Romanini, M. Barrio, J.-L. Tamarit, C. Cazorla, D. Errandonea, N. D. Mathur, A. Planes, X. Moya and L. Mañosa, Giant barocaloric effects over a wide temperature range in superionic conductor AgI, Nat. Commun. 8, 1851 (2017). Doi: https://doi.org/10.1038/s41467-017-01898-2.

27. D. Chandra, C. S. Day, C. S. Barrett, Low- and high-temperature structures of neopentylglycol plastic crystal, Powder Diffraction 8, 109 (1993). Doi: https://doi.org/10.1017/S0885715600017930.

28. M. Bée, Quasielastic Neutron Scattering Principles And Applications In Solid State Chemistry, Biology And Materials Science IOP Publishing Ltd (1988).

29. G. B. Guthrie and J. P. McCullough, Some observations on phase transformations in molecular crystals, J. Phys. Chem. Solids 18, 53 (1961). Doi: https://doi.org/10.1016/0022-3697(61)90083-X.

30. J. Font, J. Muntasell, E. Cesari and J. Pons, Solid-state mechanical alloying of plastic crystals, J. Mater. Res. 12, 3254 (1997). Doi: https://doi.org/10.1557/JMR.1997.0427.

31. F. E. Karasz and J. A. Pople, A theory of fusion of molecular crystals—II: Phase diagrams and relations with solid state transitions, J. Phys. Chem. Solids 20, 294 (1961). Doi: https://doi.org/10.1016/0022-3697(61)90017-8.

32. C.-M. Wu, G. Deng, J.S. Gardner, P. Vorderwisch, W.-H. Li, S. Yano, J.-C. Peng and E. Imamovic, SIKA—the multiplexing cold-neutron triple-axis spectrometer at ANSTO, J. Inst. 11, P10009 (2016). doi: https://doi.org/10.1088/1748-0221/11/10/P10009.

33. K. Nakajima, S. Ohira-Kawamura, T. Kikuchi, M. Nakamura, R. Kajimoto, Y. Inamura, N. Takahashi, K. Aizawa, K. Suzuya, K. Shibata, T. Nakatani, K. Soyama, R. Maruyama, H. Tanaka, W. Kambara, T. Iwahashi, Y. Itoh, T. Osakabe, S. Wakimoto, K. Kakurai, F. Maekawa, M. Harada, K. Oikawa, R. E. Lechner, F. Mezei and M. Arai, AMATERAS: A Cold-Neutron Disk Chopper Spectrometer. *J. Phys. Soc. Jpn.* **80**, SB028 (2011). doi: http://dx.doi.org/10.1143/JPSJS.80SB.SB028.

34. M. Nakamura, R. Kajimoto, Y. Inamura, F. Mizuno, M. Fujita, T. Yokoo and M. Arai, First Demonstration of Novel Method for Inelastic Neutron Scattering Measurement Utilizing Multiple Incident Energies. J. Phys. Soc. Jpn. **78**, 093002 (2009). doi: http://dx.doi.org/10.1143/JPSJ.78.093002.

35. Y. Inamura, T. Nakatani, J. Suzuki and T. Otomo, Development Status of Software "Utsusemi" for Chopper Spectrometers at MLF, J-PARC. *J. Phys. Soc. Jpn.* **82**, SA031 (2013), doi: http://dx.doi.org/10.7566/JPSJS.82SA.SA031.

36. R. T. Azuah, L. R. Kneller, Y. Qiu, P. L.W. Tregenna-Piggott, C. M. Brown, J. R. D. Copley, and R. M. Dimeo, DAVE: A comprehensive software suite for the reduction, visualization, and analysis of low energy neutron spectroscopic data, *J. Res. Natl. Inst. Stan. Technol.* **114**, 341 (2009).

37. D. Yu, R. A. Mole and G. J. Kearley, Performance test on PELICAN – a multi-purpose time of flight cold neutron spectrometer. EPJ Web of Conferences 83, 03019 (2015). doi: 10.1051/epjconf/20158303019.

38. D. Yu, R. Mole, T. Noakes, S. Kennedy, and R. Robinson Pelican — a Time of Flight Cold Neutron Polarization Analysis Spectrometer at OPAL. J. Phys. Soc. JPN 82, SA027 (2013). doi: 10.7566/jpsjs.82sa.sa027.

39. D. Richard, M. Ferrand, and G. J. Kearley, Analysis and visualisation of neutron-scattering data. J. Neutron Res. 4, 33-39 (1996). doi: 10.1080/10238169608200065.

40. G. Kresse and J. Furthmuller, Efficient iterative schemes for ab initio total-energy calculations using a plane-wave basis set. Phys. Rev. B 54, 11169 (1996). Doi: https://doi.org/10.1103/PhysRevB.54.11169.

41. G. Kresse and J. Hafner, Ab initio molecular-dynamics simulation of the liquid-metal–amorphous-semiconductor transition in germanium. Phys. Rev. B 49, 14251 (1994). Doi: https://doi.org/10.1103/PhysRevB.49.14251.

42. J. P. Perdew, K. Burke, and M. Ernzerhof, Generalized Gradient Approximation Made Simple, Phys. Rev. Lett. 77, 3865 (1996). Doi: https://doi.org/10.1103/PhysRevLett.77.3865.

43. G. Kresse and D. Joubert, From ultrasoft pseudopotentials to the projector augmented-wave method, Phys. Rev. B 59, 1758 (1999). Doi: https://doi.org/10.1103/PhysRevB.59.1758.

44. P. E. Blöchl, Projector augmented-wave method, Phys. Rev. B 50, 17953 (1994). Doi: https://doi.org/10.1103/PhysRevB.50.17953.

45. H. J. Monkhorst and J. D. Pack, Special points for Brillouin-zone integrations, Phys. Rev. B 13, 5188 (1976). Doi: https://doi.org/10.1103/PhysRevB.13.5188.

46. C. Elsässer, M. Fähnle, C. T. Chan, and K. Ho, Density-functional energies and forces with Gaussian-broadened fractional occupations, Phys. Rev. B 49, 13975 (1994). Doi: https://doi.org/10.1103/PhysRevB.49.13975.

47. S. Grimme, J. Antony, S. Ehrlich, and H. Krieg, A consistent and accurate ab initio parametrization of density functional dispersion correction (DFT-D) for the 94 elements H-Pu, J. Chem. Phys. 132, 154104 (2010). Doi: https://doi.org/10.1063/1.3382344.

48. T. Matsuo, H. Suga, W. I. F. David, R. M. Ibberson, P. Bernier, A. Zahab, C. Fabre, A. Rassat, and A.Dworkin, The heat capacity of solid C60, Solid State Commun. 83, 711-715 (1992). Doi: https://doi.org/10.1016/0038-1098(92)90149-4.